\documentclass[twocolumn,aps,floatfix,showpacs,superscriptaddress]{revtex4}
\newcommand{\be}{\begin{equation}}
\newcommand{\ee}{\end{equation}}
\newcommand{\bea}{\begin{eqnarray}}
\newcommand{\eea}{\end{eqnarray}}

\usepackage{graphicx}
\usepackage{bm}
\begin{document}

\title{$^{39}$K Bose-Einstein condensate with tunable interactions}
\author{G. Roati}
\affiliation{LENS and Dipartimento di Fisica, Universit\`a di
Firenze, INFN
  and INFM-CNR\\  Via Nello Carrara 1, 50019 Sesto Fiorentino, Italy }
  \author{M. Zaccanti}
\affiliation{LENS and Dipartimento di Fisica, Universit\`a di
Firenze, INFN
  and INFM-CNR\\  Via Nello Carrara 1, 50019 Sesto Fiorentino, Italy }
  \author{C. D'Errico}
\affiliation{LENS and Dipartimento di Fisica, Universit\`a di
Firenze, INFN
  and INFM-CNR\\  Via Nello Carrara 1, 50019 Sesto Fiorentino, Italy }
  \author{J. Catani}
\affiliation{LENS and Dipartimento di Fisica, Universit\`a di
Firenze, INFN
  and INFM-CNR\\  Via Nello Carrara 1, 50019 Sesto Fiorentino, Italy }
  \author{M. Modugno}
\affiliation{LENS and Dipartimento di Matematica Applicata,
Universit\`a di Firenze, INFN, and BEC-INFM Center, Universit\`a di
Trento, I-38050 Povo, Italy}

\author{A. Simoni}\affiliation{Laboratoire de Physique des Atomes,
Lasers, Mol\'ecules et Surfaces \\ UMR 6627 du CNRS and Universit\'e
de Rennes, 35042 Rennes Cedex, France}
\author{M. Inguscio}
\affiliation{LENS and Dipartimento di Fisica, Universit\`a di
Firenze, INFN
  and INFM-CNR\\  Via Nello Carrara 1, 50019 Sesto Fiorentino, Italy }
  \author{G. Modugno}
\affiliation{LENS and Dipartimento di Fisica, Universit\`a di
Firenze, INFN
  and INFM-CNR\\  Via Nello Carrara 1, 50019 Sesto Fiorentino, Italy }

\begin{abstract}
We produce a Bose-Einstein condensate of $^{39}$K atoms.
Condensation of this species with naturally small and negative
scattering length is achieved by a combination of sympathetic
cooling with $^{87}$Rb and direct evaporation, exploiting the
magnetic tuning of both inter- and intra-species interactions at
Feshbach resonances. We explore tunability of the self-interactions
by studying the expansion and the stability of the condensate. We
find that a $^{39}$K condensate is interesting for future
experiments requiring a weakly interacting Bose gas.

\end{abstract}
\pacs{ 03.75.-b; 34.50.-s; 32.80.Pj
 }

\date{\today}
\maketitle

Two-body interactions play a major role in the production and in the
properties of Bose-Einstein condensates (BECs) made of ultracold
atoms \cite{julienne}. Atomic species with naturally large repulsive
interactions such as $^{87}$Rb \cite{rb} or $^{23}$Na \cite{na} have
collision properties favorable for the preparation process. However,
there is growing interest in studying Bose-Einstein condensates
where the interactions can be precisely tuned, magnetic Feshbach
resonances \cite{feshbach,sodium} being a key tool in this respect.
One of the main motivations is the formation of an almost ideal
condensate, i.e. one with vanishing interatomic interactions.
Availability of such a system is essential for studying phenomena
where even a weak interaction can hide the underlying physics of
interest. A noticeable example is in the field of disordered
systems, where experiments performed with ideal quantum gases can
shed new light on the interdisciplinary phenomenon of Anderson
localization \cite{disorder,disorderexp}. An ideal BEC would also be
the most appropriate source for matter-wave interferometry,
combining maximal brightness with the absence of collisional
decoherence \cite{interferometry}. The possibility of dynamically
tuning the interactions in a BEC could also open new directions
towards Heisenberg-limited interferometry \cite{heisenberg}.

Feshbach resonances have been observed in most of the atomic species
that have been condensed so far: $^{23}$Na \cite{sodium}, $^{85}$Rb
\cite{rubidium85}, $^{133}$Cs \cite{cesium1,cesium2}, $^{7}$Li
\cite{lithium}, $^{87}$Rb \cite{rubidium87}, $^{52}$Cr
\cite{chromium}. Magnetic tuning of the interactions to small values
around zero can be performed in lithium, a possibility already
exploited to realize bright solitons in a weakly attractive BEC
\cite{lithium}. Cesium also presents an experimentally accessible
region of nearly vanishing scattering lengths at which the small
internal energy of a weakly interacting cesium BEC has been
investigated \cite{cesium2}.

In this Letter we report Bose-Einstein condensation of a new atomic
species, $^{39}$K. Combination of broad Feshbach resonances and a
small background scattering length $a_{K}\simeq$-33 $a_0$ \cite{jin}
makes this system very promising for the study of weakly interacting
condensates. Direct evaporative cooling of this species had been
prevented by unfavorable zero-field collisional properties
\cite{fort,minardi}. Sympathetic cooling with $^{87}$Rb has recently
proved \cite{minardi} to work for $^{39}$K as efficiently as for the
other potassium isotopes \cite{science,roati}, but condensation was
still prevented by the negative value of $a_{\rm K}$. We now bring
$^{39}$K to quantum degeneracy by a combination of sympathetic
cooling with $^{87}$Rb and direct evaporative cooling, exploiting
the resonant tuning of both inter- and intra-species interactions at
Feshbach resonances. Presence of one broad homonuclear Feshbach
resonance allows us to tune $a_{\rm K}$ in the condensate from large
positive values to small negative values. The possibility of
precisely adjusting $a_{\rm K}$ around zero is demonstrated by
studying the condensate expansion and its stability.

The experimental techniques we use are similar to the ones we
developed for the other potassium isotopes
\cite{science,roati,ferlaino,minardi}. We start by preparing a
mixture of $^{39}$K and $^{87}$Rb atoms in a magneto-optical trap.
The mixture contains about 10$^9$ Rb atoms at $T\simeq$100 $\mu$K
and 10$^7$ K atoms at $T\simeq$300 $\mu$K. We simultaneously load
the two species in a magnetic potential in their stretched Zeeman
states $|F=2, m_F=2\rangle$ and then we perform a selective
evaporation of rubidium on the hyperfine transition at 6.834 GHz
(the hyperfine splitting of $^{39}$K is 462 MHz). Potassium atoms
are efficiently sympathetically cooled via interspecies collision
\cite{minardi}, in spite of the small scattering length $a_{\rm
KRb}\simeq$ 28 $a_0$ \cite{ferlaino}. With this technique we are
able to prepare after 25 s of evaporation samples containing
typically 10$^6$ Rb atoms and 2$\times$10$^5$ K atoms \cite{note} at
$T$=800 nK.

Further cooling of the mixture in the magnetic potential in
principle would be possible. However, condensation would be
accompanied by collapse \cite{collapse} because of the negative
scattering length of $^{39}$K at low magnetic fields. We therefore
exploit the possibility of tuning $a_{\rm K}$ to positive values by
performing the last part of evaporation in an optical potential in
presence of a homogeneous magnetic field. We have indeed discovered
several broad Feshbach resonances for this potassium isotope
\cite{model,innsbruck}, a very favorable one being in the ground
state $|1,1\rangle$ at about 400 G. We have also performed a
detailed analysis of several of such resonances in order to
construct a quantum collisional model able to accurately predict the
magnetic-field dependence of $a_{\rm K}$ \cite{model}. We will make
use of this analysis throughout this paper.

The K-Rb mixture is adiabatically transferred to an optical trap
created with two focused laser beams at a wavelength $\lambda$=1030
nm, with beam waists of about 100$\mu$m and crossing in the
horizontal plane. The two species are then transferred to their
absolute ground states $|1,1\rangle$ via adiabatic rapid passage,
and further cooled by reducing the power in the laser beams by means
of acusto-optic modulators. The optical trap is designed in such a
way to evaporate mainly Rb. We find that cooling of $^{39}$K can be
greatly enhanced by increasing $a_{\rm KRb}$ at one of the
interspecies Feshbach resonances that exist in this mixture
\cite{ferlaino}. At a first stage lasting 2.5 s a homogeneous
magnetic field is thus tuned near a 8.5 G-wide interspecies Feshbach
resonance centered at 317.9 G \cite {note2}. We find that
sympathetic cooling is optimized at a field of 316 G, where $a_{\rm
KRb}\simeq$150 $a_0$. At this magnetic field the homonuclear
$^{39}$K cross-section is still small, $a_{\rm K}\simeq$-33 $a_0$.

\begin{figure}[htbp]
\includegraphics[width=\columnwidth,clip]{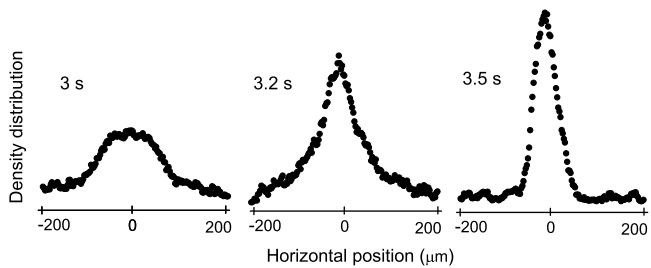}
\caption{Phase transition to a $^{39}$K BEC.  The three profiles are
 taken at different times during the final stage of forced evaporation in the optical
trap (see Fig. \ref{fig2}). The profiles are obtained by vertically
integrating the column density measured after release from the trap
and 15 ms of ballistic expansion. } \label{fig1}
\end{figure}

\begin{figure}[htbp]
\includegraphics[width=\columnwidth,clip]{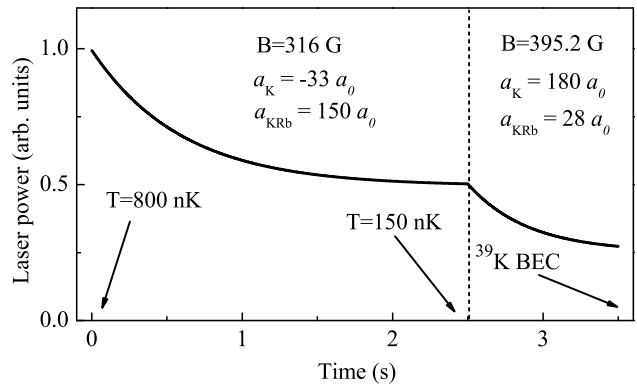}
\caption{Sketch of the relevant parameters of the system during the
evaporation in the optical potentials. The two exponential ramps
have time constants of 0.9 s and 0.45 s, respectively.} \label{fig2}
\end{figure}

When both gases are close to quantum degeneracy ($T\simeq$150 nK) we
make $a_{\rm K}$ positive and large by shifting the magnetic field
in proximity of a 52 G-wide $^{39}$K resonance, centered at 402.4 G,
and we continue the evaporation for 1 s. Due to the different trap
depths for the two species, Rb is soon completely evaporated and
further cooling of K relies just on intra-species collisions. We
find for this phase an optimal scattering length $a_{\rm
K}\simeq$180 $a_0$ obtained for $B$=395.2 G. At this field the two
species are only weakly coupled, since $a_{\rm KRb}\simeq$28 $a_0$.
Fig. \ref{fig1} shows the phase transition of the K cloud to a
Bose-Einstein condensate, detected via absorption imaging after a
ballistic expansion. The critical temperature we measure is around
100 nK, and our purest condensates contain typically 3$\times$10$^4$
atoms. The frequencies of the optical trap at the end of the
evaporation are $\omega$=2$\pi \times$ (65, 74, 92) s$^{-1}$ in the
($x$, $y$, $z$) directions respectively.  The whole evaporation
procedure in the optical trap is summarized in Fig. \ref{fig2}.

\begin{figure}
\includegraphics[width=\columnwidth,clip]{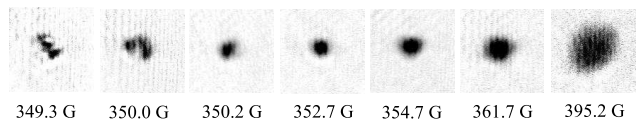}
\caption{Sample images of a $^{39}$K condensate at various magnetic
fields in the vicinity of the 52 G-wide Feshbach resonance centered
at 402.4 G, taken after 31.5 ms of expansion. The field of view is
300$\times$300 $\mu$m. The size shrinks as the scattering length
$a_{\rm K}$ is decreased, and the condensate eventually collapses
for negative $a_{\rm K}$.} \label{fig5}
\end{figure}

\begin{figure}
\includegraphics[width=\columnwidth,clip]{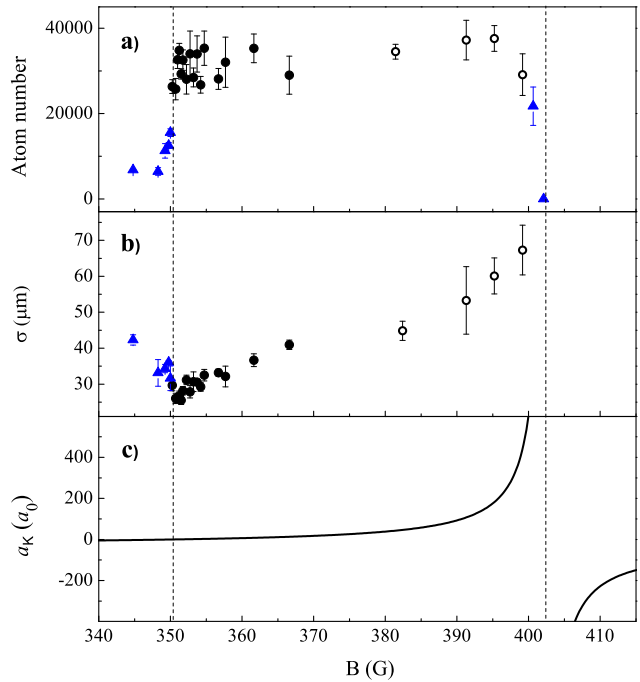}
\caption{(color online) Tuning the interaction in a $^{39}$K
condensate. a) atom number; b) width of the cloud after 31.5 ms of
ballistic expansion; c) theory prediction for the scattering length.
The two dashed lines indicate the expected position of the
zero-crossing and resonance center. Condensates are either fitted
with a Thomas-Fermi profile (circles) in the region of large
interactions or a gaussian profile (dots) in the region of weak
interactions. Atom number and width of uncondensed clouds are
directly extracted from the raw images (triangles). Each data point
is the average of at least three measurements.} \label{fig3}
\end{figure}

Once the condensate is produced, $a_{\rm K}$ can be further tuned.
We have explored the magnetic-field region below the homonuclear
Feshbach resonance in which the condensate is stable. The experiment
starts with a pure BEC created at $B_0$=395.2 G. The field is then
adiabatically brought to a final field $B$ in 30 ms. After 5 ms the
optical trap is switched off and the cloud expands for 31.5 ms
before absorption imaging is performed with a resonant beam
propagating along the $y$ direction. The magnetic field is switched
off just 5 ms before imaging, to ensure that $a_{\rm K}$ does not
change during the relevant phases of the expansion. Examples of
absorption images are shown in Fig. \ref{fig5}.  The measured atom
number and the mean width $\sigma$=$(\langle x^2\rangle$ + $\langle
z^2\rangle)^{1/2}$ are shown Fig. \ref{fig3}, together with the
magnetic-field dependence of $a_{\rm K}$ as calculated using our
quantum collision model \cite{model}.

Between 350.2 G and 350.0 G we observe a sudden drop of the atom
number that can be attributed to a collapse \cite{collapse} of the
BEC for too large negative $a_{\rm K}$. In this regime the sample is
no more in equilibrium and the presence of strong excitations is
evident, see leftmost panels in Fig. \ref{fig5}. On the other
extreme, the field can be brought in proximity of the resonance
center. Here we observe that the BEC is depleted by three-body
recombination, whose rate scales as $a_K^4$ \cite{threebody} close
to a Feshbach resonance. For example, the lifetime of the BEC in the
optical trap, which is typically around 3 s, is shortened to about
200 ms when the field is set to 399.2 G, where $a_{\rm K}$=440 (40)
G.

The width of the condensate after the expansion shown in Fig.
\ref{fig3}b features a decrease by almost a factor three as the
field strength shifts from the resonance to the zero-crossing
region. This is due to the variation of the interaction strength in
the condensate. At the long expansion time of this experiment
$\sigma\propto\sqrt{E_{rel}}$, where $E_{rel}$ is the release energy
of the condensate. The latter quantity is expected to decrease as
$a_K^{2/5}$ in the Thomas-Fermi limit, i.e. for large positive
$a_{\rm K}$. Its value equals the kinetic energy of the harmonic
oscillator ground state for $a_{\rm K}$=0, and becomes even smaller
for $a_K<0$. The sharp increase of the width below 350.2 G reported
in Fig. \ref{fig3} reflects the presence of excitations in the
collapsed system.

\begin{figure}[htbp]
\includegraphics[width=\columnwidth,clip]{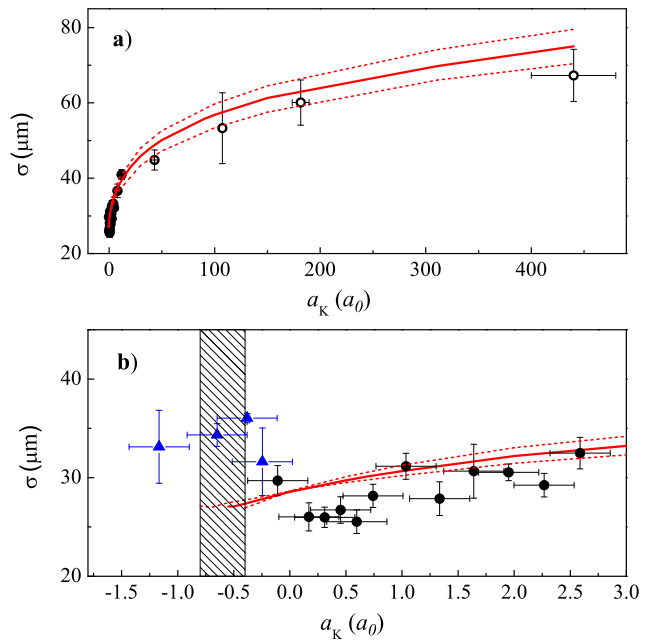}
\caption{(color online) a) Scattering-length dependence of measured
(dots) and calculated (lines) mean width $\sigma$. The two dashed
lines indicate the range of variation of $\sigma$ due to a 30\%
systematic uncertainty on atom numbers. b) Zoom into the
zero-crossing region. Both condensed samples (dots) and collapsed
samples (triangles) are shown. The horizontal error bars are
determined by the uncertainty on $a_K(B)$. The hatched region
indicates the critical $a_{\rm K}$ for collapse as predicted by the
numerical calculation, including its variation due to the
uncertainty in the atom number.} \label{fig4}
\end{figure}

To gain insight into the observed phenomenology, we have compared
the condensate widths with numerical calculations based on the
Gross-Pitaevskii theory \cite{stringari}. In Fig. \ref{fig4} we plot
$\sigma$ as a function of $a_{\rm K}$; here the abscissa values for
the experimental data have been calculated using the theoretical
$a_K(B)$ \cite{model}. The horizontal error bar is dominated by the
uncertainty in the model for $a_K(B)$, which amounts to about 0.27
$a_0$ in the zero-crossing region. The decrease in $\sigma$ with
decreasing $a_{\rm K}$ is the result of two general effects: i) a
reduction of the condensate width in the trap; ii) a reduction of
the interaction energy released during the first phases of the
expansion. Note in Fig. \ref{fig4}a the good agreement between
theory and experiment in the broad range of values of $a_{\rm K}$ in
which the condensate is stable. The slow decrease of $\sigma$ for
moderately large and positive $a_{\rm K}$ is followed by a faster
decrease in the region of the zero-crossing.

In Fig. \ref{fig4}b we compare theory and experiment on a much
smaller region around the zero-crossing, including also the
experimental data points corresponding to a collapsed cloud. The
hatched region indicates the critical scattering length for collapse
$a_c$=-0.57(20) $a_0$ predicted by the theory for the nominal atom
number we had in this experiment, $N$=3.5$\times$10$^4$. The width
of the cloud keeps on decreasing as $a_{\rm K}$ gets negative and
increases again at collapse. In this experiment collapse is
apparently happening at a slightly subcritical scattering length
$a_{\rm K}$=-0.2(3) $a_0$. This might be due to a loss of
adiabaticity of our magnetic-field ramp in this region of negative
$a_{\rm K}$. Although the ramp duration is much longer than the trap
period, it might still excite the monopole collective mode of the
condensate which has a vanishing frequency for $a_{\rm K}$
approaching $a_c$ \cite{stringarib}.

In conclusion, we have produced a Bose-Einstein condensate of
$^{39}$K atoms in which the scattering length can be precisely tuned
over a large range and adjusted around zero. This atomic species is
particularly advantageous in producing a weakly interacting
condensate, since it combines a broad Feshbach resonance with a
small background scattering length. For this resonance the
theoretical model \cite{model} predicts a sensitivity $da_{\rm
K}$/$dB \simeq$ 0.55 $a_0$/G around 350 G. Therefore, a
magnetic-field stability of the order of 0.1 G will in principle
allow us to tune the scattering length to zero to better than 0.1
$a_0$ in future experiments. This degree of control appears superior
to that achievable in most other species which present either
narrower resonances and/or larger background scattering lengths, the
only exception being $^7$Li \cite{lithium}.

A weakly interacting $^{39}$K Bose gas may have a variety of
applications, ranging from Anderson localization of matter-waves to
high-sensitivity atom interferometry. We note that the in-trap size
of a the weakly-interacting condensate is comparable to that of the
ground state of the trapping potential, that is
$\sqrt{\hbar/m\overline{\omega}}$ = 1.84 $\mu$m in the present
experiment. Such small size opens interesting perspectives for
inertial measurements with high spatial resolution \cite{carusotto}.

We expect that also a binary $^{39}$K-$^{87}$Rb BEC can be
efficiently produced, thus enriching the possibilities offered by
potassium-rubidium mixtures. Such binary condensate could be used
for various applications and is especially appealing for the
production of ultracold heteronuclear molecules.

We are indebted to F. Minardi and L. De Sarlo for precious
suggestions. We thank M. Fattori, F. Ferlaino, F. Marin and all the
other members of the Quantum Gases group at LENS for contributions
and useful discussions, and M. De Pas, M. Giuntini, A. Montori, R.
Ballerini, and A. Hajeb for technical assistance. This work was
supported by MIUR, by EU under contracts HPRICT1999-00111 and
MEIF-CT-2004-009939, by INFN, by Ente CRF, Firenze and by CNISM,
Progetti di Innesco 2005.


\begin{thebibliography}{99}
\bibitem{note} The decrease of the potassium atom number is
due to spin-exchange inelastic collisions with Rb atoms in states
other than $|2,2\rangle$.
\bibitem{note2} We find this resonance at a magnetic field very close to the
value predicted in \cite{ferlaino}. A detailed analysis of Feshbach
resonances in this mixture will be the subject of a future
publication.
\end{thebibliography}
\end{document}